\documentclass[
  twocolumn,
  aps,prl,showpacs,amsmath,
  amssymb,superscriptaddress,
  bibnotes,
  ]{revtex4-1}

\usepackage{hyperref}
\usepackage{graphicx}
\usepackage{mathtools}
\usepackage{hyperref}
\usepackage{color}
\usepackage[svgnames]{xcolor}
\usepackage{float}
\usepackage{lipsum}

\usepackage{changes}

\usepackage[export]{adjustbox}

\newcommand{\chiunit}{\,\mu_B^2/\textrm{eV}}
\newcommand{\eV}{\,\textrm{eV}}
\newcommand{\Q}{\mathbf{Q}}
\newcommand{\QIC}{\mathbf{Q}_{SDW}}
\newcommand{\TQIC}{\tilde{\mathbf{Q}}_{SDW}}
\newcommand{\QX}{\mathbf{Q}_{X}}
\newcommand{\QM}{\mathbf{Q}_{M}}
\newcommand{\QG}{\mathbf{Q}_{\Gamma}}

\newif\ifshowcomments\showcommentstrue
\begin{document}
% --------------------------------------------------------------------

% --------------------------------------------------------------------
\title{Magnetic response of Sr$_2$RuO$_4$: quasi-local spin fluctuations due to Hund's coupling}
% --------------------------------------------------------------------

% --------------------------------------------------------------------
\author{Hugo U.~R.~Strand}
\email{hugo.strand@gmail.com}
\affiliation{Center for Computational Quantum Physics, Flatiron institute, Simons Foundation, 162 5th Ave., New York, 10010 NY, USA} 

\author{Manuel Zingl}
\affiliation{Center for Computational Quantum Physics, Flatiron institute, Simons Foundation, 162 5th Ave., New York, 10010 NY, USA} 

\author{Nils Wentzell}
\affiliation{Center for Computational Quantum Physics, Flatiron institute, Simons Foundation, 162 5th Ave., New York, 10010 NY, USA} 

\author{Olivier Parcollet}
\affiliation{Center for Computational Quantum Physics, Flatiron institute, Simons Foundation, 162 5th Ave., New York, 10010 NY, USA}
\affiliation{Institut de Physique Th\'eorique (IPhT), CEA, CNRS, UMR 3681, 91191 Gif-sur-Yvette, France}

\author{Antoine Georges}
\affiliation{Center for Computational Quantum Physics, Flatiron institute, Simons Foundation, 162 5th Ave., New York, 10010 NY, USA} 
\affiliation{Coll\`ege de France, 11 place Marcelin Berthelot, 75005 Paris, France}
\affiliation{Centre de Physique Th\'eorique, Ecole Polytechnique, CNRS, 91128 Palaiseau Cedex, France}
\affiliation{Department of Quantum Matter Physics, University of Geneva, 24 Quai Ernest-Ansermet, 1211 Geneva 4, Switzerland}

\date{\today} 
%\pacs{}
% --------------------------------------------------------------------

% --------------------------------------------------------------------
\begin{abstract}
We study the magnetic susceptibility in the normal state of Sr$_2$RuO$_4$ using dynamical mean-field theory including dynamical vertex corrections.
Besides the well known incommensurate response, our calculations yield quasi-local spin fluctuations which are broad in momentum and centered around the $\Gamma$ point, in agreement with recent  inelastic neutron scattering experiments [P. Steffens, \textit{et al.}, Phys. Rev. Lett. \textbf{122}, 047004 (2019)].
We show that these quasi-local fluctuations are controlled by the Hund's coupling and account for the dominant contribution to the momentum-integrated response.
While all orbitals contribute equally to the incommensurate response, the enhanced $\Gamma$ point response originates from the planar xy orbital.
\end{abstract}
% --------------------------------------------------------------------

\maketitle
\makeatletter
\let\toc@pre\relax
\let\toc@post\relax
\makeatother

% --------------------------------------------------------------------
% Introduction
% --------------------------------------------------------------------

The importance of spin fluctuations for the physics of Sr$_2$RuO$_4$ has been
emphasized long ago \cite{PhysRevLett.83.3320}. This material is close to a
spin-density-wave instability and small concentrations of impurities 
trigger ordering \cite{PhysRevB.63.180504, PhysRevLett.88.197002}. 
Inelastic neutron scattering experiments (INS) pioneered by Sidis \emph{et al.\
}\cite{PhysRevLett.83.3320} and refined over the years
\cite{PhysRevB.65.184511, PhysRevB.66.064522, Braden:2003aa,
   PhysRevLett.92.097402, PhysRevB.84.060402, doi:10.1143/JPSJ.81.124710,
   PhysRevLett.122.047004} have revealed that the magnetic response is
essentially the sum of 
(i) a weakly momentum-dependent contribution centered at
$\Gamma$ (in agreement with the Stoner enhancement factor
of the uniform susceptibility by $\sim 7$ as compared to the band value
\cite{0953-8984-7-47-002, RevModPhys.75.657})
and 
(ii) a peak at an incommensurate wavevector $\QIC \approx (0.3,
0.3, 0)$ \footnote{In units of the reciprocal tetragonal lattice vectors $2\pi/a$ and $2\pi/c$.} signaling the proximity to a spin-density-wave (SDW) instability
\cite{PhysRevLett.122.047004}. 
The peak at $\QIC$ was predicted by Mazin and Singh \cite{PhysRevLett.82.4324}
using density functional theory  (DFT) and  the random phase approximation (RPA).  
However, DFT+RPA does not account for the broad structure at $\Gamma$, and it also predicts substantial anti-ferromagnetic fluctuations at the $X$ point, $\QX \!=\! (0.5, 0.5, 0)$, in contradiction to experiments \cite{PhysRevLett.122.047004}.

More recently, however, it has been realized that the origin of the strong correlations in this material 
may not be associated with long-wavelength magnetic correlations, 
but with local correlations driven by the Hund's coupling \cite{PhysRevLett.106.096401, Georges:2013fk}.
A successful description of an extensive set of physical properties of Sr$_2$RuO$_4$ has
been obtained following this picture, supported by quantitative dynamical mean-field (DMFT) calculations.  
This includes the large mass enhancements of quasiparticles observed in de Haas-van Alphen experiments
\cite{PhysRevLett.76.3786, *PhysRevLett.84.2662,
   *doi:10.1080/00018730310001621737} and angle resolved photo-emission
spectroscopy \cite{Mackenzie:1998aa, *PhysRevLett.85.5194, *PhysRevB.72.104514,
   *PhysRevB.72.205114, *PhysRevLett.105.226406, *PhysRevLett.109.066404,
   *Zabolotnyy:2013aa, *PhysRevLett.116.197003}, as well as quasi-particle weights and lifetimes
\cite{PhysRevLett.106.096401}, nuclear magnetic resonance
\cite{PhysRevLett.106.096401}, optical conductivity
\cite{PhysRevLett.113.087404, PhysRevLett.116.256401}, thermopower
\cite{PhysRevLett.117.036401}, Hall coefficient \cite{2019arXiv190205503Z},
quasiparticle dispersions \cite{PhysRevLett.116.106402, PhysRevLett.120.126401,
   2018arXiv181206531T}, and magnetic response \cite{Boehnke:2014fk,
   0295-5075-122-5-57001, Gingras:2018aa, 2018arXiv181105143A}.  

% --------------------------------------------------------------------
\begin{figure}[b]
\includegraphics[scale=1.0]{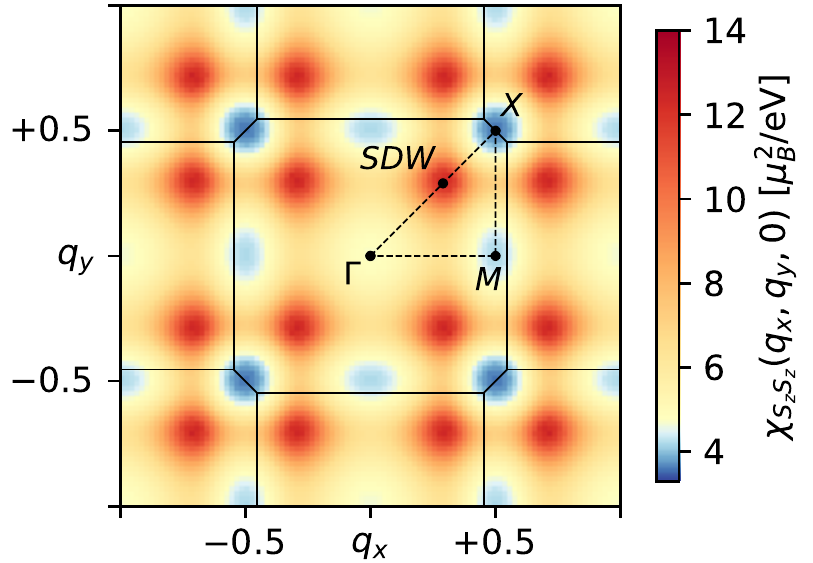}
\caption{\label{fig:plane}
  Spin-susceptibility $\chi_{S_z S_z}(\mathbf{Q})$ from DMFT at $T=464\,$K in the $q_x, q_y$-plane at $q_z = 0$, with incommensurate hot-spots at $\QIC$ (red), cold-spots at $\QM$ and $\QX$ (blue), and a broad response centered around $\QG$ (yellow), in units of the reciprocal tetragonal lattice vectors $2\pi/a$ and $2\pi/c$.
} \end{figure}
% --------------------------------------------------------------------

In this letter, we bridge this gap between the spin fluctuations picture and the
Hund's metal picture of the normal state of Sr$_2$RuO$_4$ 
by analyzing the magnetic response function using DMFT. 
Our results reproduce the overall momentum dependence obtained in experiments \cite{PhysRevLett.122.047004}, see Fig.~\ref{fig:plane}, and reveal strong coupling effects which cannot be accounted for in RPA, such as a suppression of the antiferromagnetic response at $\QX$.
We find that the response is dominated by quasi-local (weakly momentum-dependent) spin fluctuations,
and show that these fluctuations are controlled by the strength of the Hund's coupling.
As discussed at the end of this letter, our findings have direct relevance for theories of the superconducting pairing mechanism, which is still an outstanding and much debated question~\cite{Mackenzie:2017aa}. 

% --------------------------------------------------------------------
%\para{Method:}
% --------------------------------------------------------------------
%
We compute the magnetic susceptibility $\chi_{S_z S_z}(\mathbf{Q})$ using DMFT
\cite{Georges:1996aa}, a DFT derived effective three-band $t_{2g}$ model without
spin-orbit coupling \footnote{The model is identical to the one used in \cite{2018arXiv181206531T} and derived from a density functional theory (DFT) calculation of Sr$_2$RuO$_4$ in the experimental structure \cite{PhysRevB.52.R9843} (at $100\,$K) using Wien2k \cite{wien2kBook} with
PBE \cite{PhysRevLett.77.3865} and $20^3$ k-points. Maximally-localized
$t_{2g}$ Wannier functions were constructed using wien2wannier
\cite{Kunes:2010ve} and Wannier90 \cite{PhysRevB.56.12847, Mostofi:2008aa,RevModPhys.84.1419} (using $10^3$ k-points).}, and a local Kanamori
interaction \cite{Kanamori:1963aa} with Hubbard $U=2.3\eV$ and Hund's
$J=0.4\eV$ \cite{PhysRevLett.106.096401}.
The DMFT equations were solved using the hybridization expansion continuous time quantum Monte Carlo \cite{Werner:2006rt, *Werner:2006qy, *Haule:2007ys, *Gull:2011lr} implementation in TRIQS \cite{Parcollet2015398, Seth2016274}. The DMFT particle-hole irreducible vertex was used to compute the static lattice susceptibility from the Bethe-Salpeter equation (BSE) \cite{Georges:1996aa} as implemented in the TRIQS two-particle response function toolbox \cite{Strand:tprf}.
Moreover, the static response at three specific momenta was computed down to much lower temperature, using self-consistent DMFT in applied magnetic fields by zero field extrapolation in supercells.
%

% --------------------------------------------------------------------
%\para{DMFT momentum structure:}
% --------------------------------------------------------------------
%
Fig.~\ref{fig:plane} displays the momentum dependent magnetic susceptibility from DMFT, with hot-spots at $\QIC$ and ridges in $q_x$ and $q_y$ connecting these hot spots.
This SDW component can be understood from the DFT
electronic structure \cite{PhysRevB.51.1385, *PhysRevB.52.1358, *Hase:1996aa,
   *McMullan:1996aa, *PhysRevB.59.9894, *PhysRevB.74.035115} of this material,
which has three Ru(4$d$)-$t_{2g}$ bands crossing the Fermi level, filled with
four electrons. 
The quasi-two-dimensional $\gamma$ band, with dominant xy orbital content, has
a larger bandwidth (by a factor of $\sim\!2$) and slightly lower energy than
the quasi-one-dimensional $\alpha$ and $\beta$ bands, originating mainly from
the xz and yz orbitals.
The peak at $\QIC$ is generated by nesting in the $\alpha$ and $\beta$ (xz, yz) Fermi
surface sheets, yielding ridges at $(0.3, q_y, 0)$ and $(q_x, 0.3, 0)$ that
cross and produce the peak at $\QIC$  \cite{PhysRevLett.82.4324}. 

The response in Fig.~\ref{fig:plane} also shows a large component, broad in momentum,
with enhanced intensity centered at $\Gamma$ in comparison to the cold-spots at $M$ and $X$. 
This is the signature of the important quasi-local spin fluctuations.
Antiferromagnetic fluctuations are suppressed, with the $X$ point being the global minimum of the response. 
This is qualitatively different from the results of weak-coupling approaches such as RPA \cite{PhysRevLett.82.4324, PhysRevLett.86.5978, doi:10.1143/JPSJ.69.3764, PhysRevB.65.220502, 0295-5075-58-6-871} -- even when basing RPA on the dressed DMFT Lindhard function~\cite{Gingras:2018aa} -- or the fluctuation-exchange approximation \cite{PhysRevB.90.245103, PhysRevB.91.159906}.
In  contrast, these approximations yield an enhanced response at the $X$ point and fail to account for the quasi-local response. The latter was not discussed in previous DMFT work \cite{0295-5075-122-5-57001}, but noted in a recent DMFT+GW calculation \cite{2018arXiv181105143A}. Both the quasi-local response and the suppression of the $X$-point fluctuations are in qualitative agreement with the recent INS experiments \cite{PhysRevLett.122.047004}. 

% --------------------------------------------------------------------
\begin{figure}[t]
\includegraphics[scale=1.0]{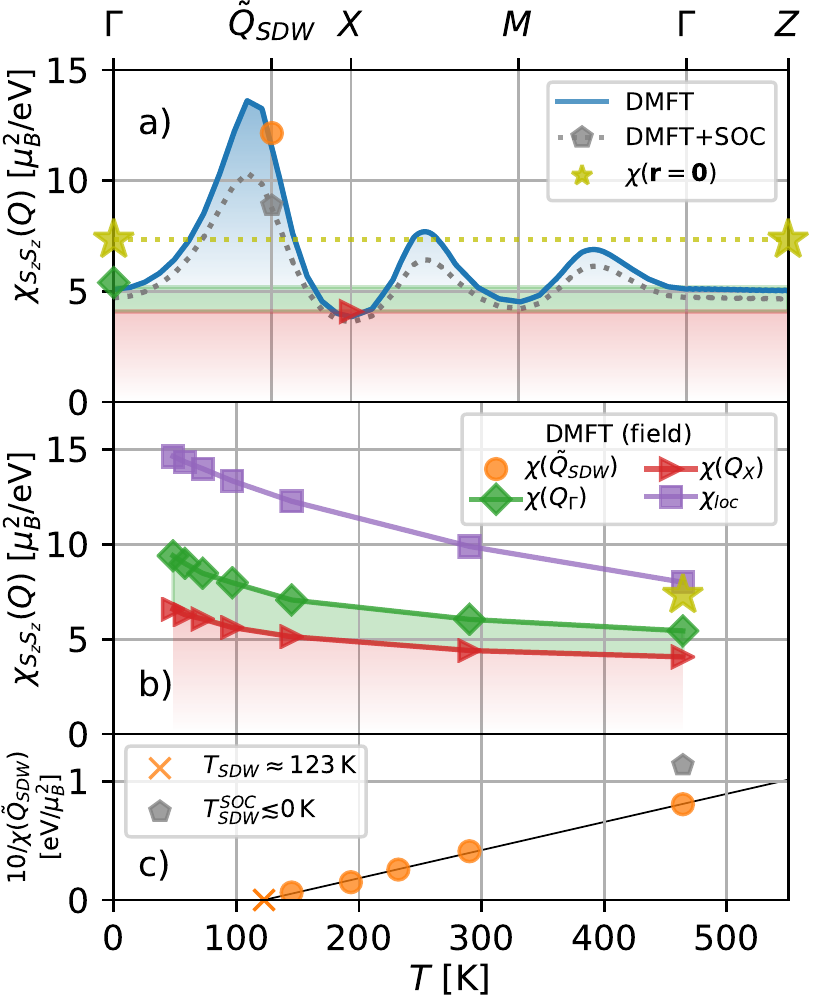}
\caption{\label{fig:path}
  a) Spin-susceptibility $\chi_{S_z S_z}(\mathbf{Q})$ from DMFT at $T=464\,$K, along the high-symmetry path $\Gamma$--$X$--$M$--$\Gamma$--$Z$ (see Fig.\ \ref{fig:plane}) with (gray dotted line) and without (blue line) SOC, together with the applied field response at $\QG$ (green diamond), $\QX$ (red triangle), $\TQIC$ (orange circle), and $\chi(\mathbf{r} \! = \! \mathbf{0})$ (yellow star).
  b) Temperature dependence of $\chi(\QX)$, $\chi(\QG)$, and $\chi_{loc}$.
  c) Temperature dependence of $1/\chi(\TQIC)$ without (circles) and with (pentagon) SOC.
} \end{figure}
% --------------------------------------------------------------------

% --------------------------------------------------------------------
%\para{High-symmetry path:}
% --------------------------------------------------------------------
%
Studying the susceptibility along the high symmetry path $\Gamma$--$X$--$M$--$\Gamma$--$Z$ gives a quantitative picture of the response, see Fig.\ \ref{fig:path}a.
The incommensurate response at $\QIC$ yields a peak on $\Gamma$--$X$ and the nesting ridges become local maxima on $X$--$M$ and $M$--$\Gamma$.
The response at $\QG = (0, 0, 0)$ is enhanced relative to the cold-spots at $\QX = (0.5, 0.5, 0)$ and $\QM = (0.5, 0, 0)$ (green-shaded area) with $\QX$ being the global minimum.
We note in passing that the negligible dispersion on $\Gamma$--$Z$ shows that the response is quasi two-dimensional. 
The quasi-local response (red- and green-shaded area) is the dominant part of the susceptibility, accounting for more than half of the momentum averaged response (yellow stars).

% --------------------------------------------------------------------
%\para{Thermodynamical consistency:}
% --------------------------------------------------------------------
%
We also perform complementary calculations of the susceptibility down to much lower temperature through self-consistent DMFT in applied fields at $\QG$, $\QX$, and in the vicinity of the incommensurate wave vector $\QIC$ at $\TQIC=(1/3, 1/3, 0)$ (using a $\sqrt{2}\times\sqrt{5}$ three site supercell), see Fig.~\ref{fig:path}b and \ref{fig:path}c.
The result is in quantitative agreement with the DMFT response obtained from the BSE after extrapolating to infinite fermionic frequency cutoff \cite{Luitz2013}, see markers in Fig.~\ref{fig:path}a.
This serves as a non-trivial consistency check of our calculations and is, to the best of our knowledge, the first demonstration of thermodynamical consistency in DMFT at the two-particle level in a multiorbital model \cite{Georges:1996aa, Potthoff:2006aa, PhysRevB.90.235105}.

% --------------------------------------------------------------------
%\para{Temperature dependence:}
% --------------------------------------------------------------------
%
When lowering temperature the spin susceptibility is enhanced, see Fig.\ \ref{fig:path}b. In particular, both $\chi(\QG)$ and $\chi(\QX)$ grow with decreasing temperature, where $\chi(\QX)$ can be taken as a direct measure of the background response (red shaded area).
However, the relative $\Gamma$ point enhancement (green shaded area) is robust and roughly constant, $\chi(\QG)/\chi(\QX) \approx 4/3$, in the studied temperature range. The precise value of this ratio, however, strongly depends on $J$ (see below).
The DMFT local impurity susceptibility $\chi_{loc}$ shows a similar temperature dependence, and is approximately equal to the local susceptibility $\chi(\mathbf{r} = \mathbf{0}) \equiv \frac{1}{V} \sum_{\mathbf{Q}} \chi(\mathbf{Q})$ at $T=464\,$K.
This rough agreement strengthens the use of $\chi_{loc}$ as a proxy for the momentum average, $\chi_{loc} \sim \chi(\mathbf{r} = \mathbf{0})$ \cite{PhysRevLett.106.096401, PhysRevLett.117.036401}.

% --------------------------------------------------------------------
%\para{Magnetic order:}
% --------------------------------------------------------------------
%
While it is known that pristine Sr$_2$RuO$_4$ does not order magnetically \cite{RevModPhys.75.657}, 
the question of whether DMFT yields SDW order at low temperature (like DFT \cite{Kim:2017aa}) has not been addressed previously.
To answer this question we make a linear extrapolation of $\chi^{-1}(\TQIC)$ in temperature, see Fig.~\ref{fig:path}c. For the established values of $U$ and $J$ \cite{PhysRevLett.106.096401}, and in the absence of spin-orbit coupling, we find that DMFT yields SDW order at $T_{SDW} \approx 123$~K, much lower than RPA \footnote{This is a drastic suppression compared to RPA with bare interactions on the DMFT bubble with $T_{SDW} \sim 2500\,$K and the Hartree-Fock bubble with $T_{SDW} \sim 6850\,$K.}.
However, the transition temperature is very sensitive to the precise value of the microscopic parameters, in particular the Hund's coupling $J$ (not shown). 

The sensitivity in $J$ raises the question how the relatively small spin-orbit coupling (SOC) $\lambda_{DFT} \approx 0.1\eV$ \cite{PhysRevLett.116.106402} affects the ordering temperature.
Full DMFT calculations with SOC, in the relevant temperature range, are out of reach with currently available algorithms.
Instead we resort to an approximate treatment -- following Ref.\ \onlinecite{PhysRevLett.120.126401} -- and add a static self-energy correction to the DMFT bubble in the BSE, with a correlation-enhanced SOC coupling $\lambda = 2\lambda_{DFT}$, see also Ref.\ \onlinecite{PhysRevLett.101.026408}. This accounts for the first order SOC contributions to the DMFT bubble $\chi^{(0)}$ but neglects the effect of SOC on the vertex.
In momentum space the magnetic susceptibility with SOC corrections exhibits an overall suppression of the incommensurate and ridge response while the $\Gamma$, $X$, and $M$ points are only weakly affected, see Fig.\ \ref{fig:path}a.
The reduced incommensurate response yields a higher inverse susceptibility, see Fig.\ \ref{fig:path}c, shifting the transition to lower temperature. Using the linear slope of the non-SOC case gives $T^{SOC}_{SDW} \lesssim 0$.
Our tentative conclusion is thus that a full DMFT+SOC calculation down to low temperature is likely not to yield SDW ordering. This obviously deserves further studies.  
Since the inclusion of SOC primarily affects the SDW response, which is not the main focus of our study, we will neglect it in the following.

% --------------------------------------------------------------------
\begin{figure}[t]
\includegraphics[scale=1.00]{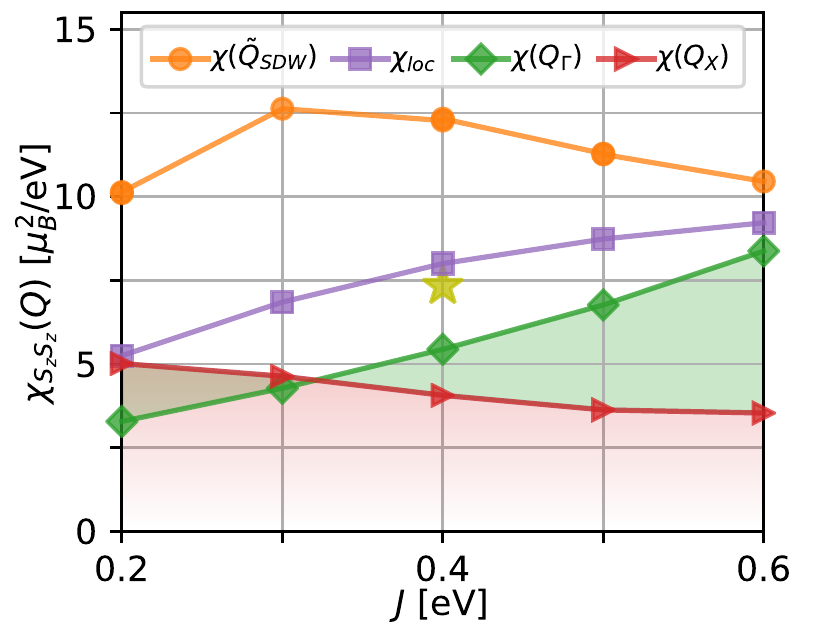}
\caption{\label{fig:sweep}
  Spin-susceptibility $\chi_{S_z S_z}(\mathbf{Q})$ for $T = 464\,$K from DMFT at $\TQIC$ (orange circles), $\QG$ (green diamonds), $\QX$ (red triangles), and the impurity local susceptibility $\chi_{loc}$ (purple squares), as a function of $J$ at $U = 2.3\eV$ and around $J=0.4\eV$. The local lattice susceptibility $\chi(\mathbf{r}\!\! =\!\! \mathbf{0})$ (yellow star) is also shown. 
} \end{figure}
% --------------------------------------------------------------------

% --------------------------------------------------------------------
%\para{Signatures of Hund's coupling:}
% --------------------------------------------------------------------
%
To disentangle the microscopic mechanisms driving the different components of the magnetic response, we study their dependence on the Hund's coupling $J$, see Fig.~\ref{fig:sweep}.
While the incommensurate spin-density-wave response $\chi(\TQIC)$ displays a non-monotonic behavior in $J$, increasing $J$ suppresses $\chi(\QX)$ and drastically increases $\chi(\QG)$ and $\chi_{loc}$.
Hence, the Hund's coupling drives the observed $\Gamma$ point enhancement (green shaded area in Figs.\ \ref{fig:path}a and \ref{fig:path}b), as well as the enhancement of the local susceptibility. Since the response around the $\Gamma$ point is very broad in momentum space (see Fig.\ \ref{fig:plane}) this in turn suggests that the Hund's coupling is responsible for the overall quasi-local magnetic response.
We note in passing that the opposite trends of $\chi(\QG)$ and $\chi(\QX)$ as a function of $J$ produces a qualitative change of the magnetic response at $J \sim 0.32\eV$ where the two terms cross.
We conclude that the Hund's coupling is responsible for the enhanced quasi-local fluctuations and plays a key role in the overall momentum space structure of the magnetic response.

\begin{figure}[t]
%  \ \\[-20mm]
\includegraphics[scale=1.0]{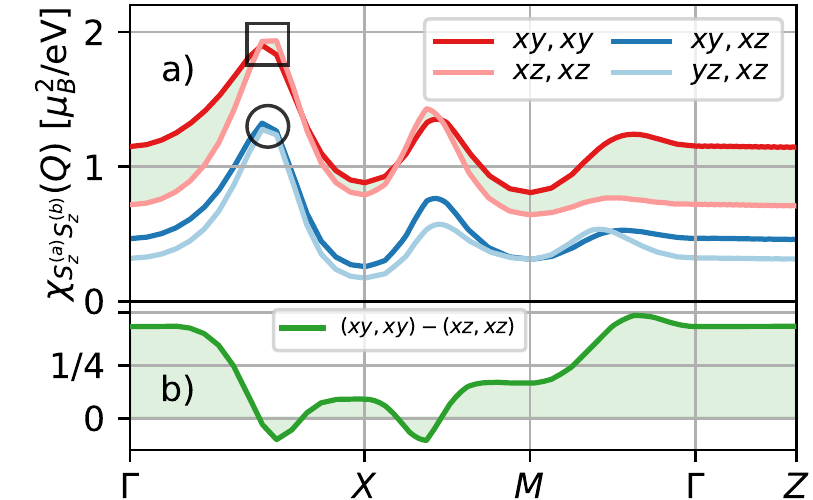}
\caption{\label{fig:orb_bse} 
a) Orbitally resolved $\chi_{S_z^{(a)} S_z^{(b)}}(\mathbf{Q})$ with $a,b \in \{xy, xz, yz\}$ at $T=464\,$K from DMFT with diagonal $xy,xy$ (red) and $xz,xz$ (light red) response and off-diagonal $xy,xz$ (blue) and $yz,xz$ response (light blue) which contribute equally at $\QIC$ (black markers).
b) The difference in the diagonal orbital response $(xy,xy) - (xz, xz)$ (green).
} \end{figure}

% --------------------------------------------------------------------
%\para{Orbital decomposition:}
% --------------------------------------------------------------------
%
We finally investigate how the magnetic response is distributed over the planar $xy$, and out-of-plane $xz$ and $yz$ orbitals, by studying the decomposition
\begin{equation*}
  \chi(\Q) \! \equiv \chi_{S_z S_z}(\Q) \! = \! \sum_{ab} \chi_{S_z^{(a)} S_z^{(b)}}(\Q)
  \, , \,\,\,
  a,b \in \{ xy, xz, yz \}
  \, ,
\end{equation*}
shown in Fig.~\ref{fig:orb_bse}a. We find that the orbital-off-diagonal response ($a \ne b$) is roughly 50\% of the total magnetic response and confirm \cite{0295-5075-122-5-57001} that $xy$, $xz$, and $yz$ contribute approximately equally to the $\QIC$ response, see markers in Fig.~\ref{fig:orb_bse}.
However, $\chi_{S_z^{(xy)} S_z^{(xy)}}(\Q)$ is markedly higher than $\chi_{S_z^{(xz)} S_z^{(xz)}}(\Q)$ around $\Gamma$ and along $M \!-\! \Gamma$.
It is this part of the $\chi_{S_z^{(xy)} S_z^{(xy)}}(\Q)$ response, shown in Fig.~\ref{fig:orb_bse}b, that is the origin of the broad plateau around $\Gamma$ and cold spots at $X$ and $M$ in Fig.~\ref{fig:plane}, and the $\QG$ enhancement (green area) in Fig.~\ref{fig:path}a.
While this only gives a weak momentum dependence to the large quasi-local magnetic response (red and green areas in Fig.~\ref{fig:path}a), the momentum space variations are extremely sensitive to the Hund's coupling, as seen in Fig.~\ref{fig:sweep}.

% --------------------------------------------------------------------
% --------------------------------------------------------------------

\begin{figure}[t] 
\includegraphics[scale=1.0] {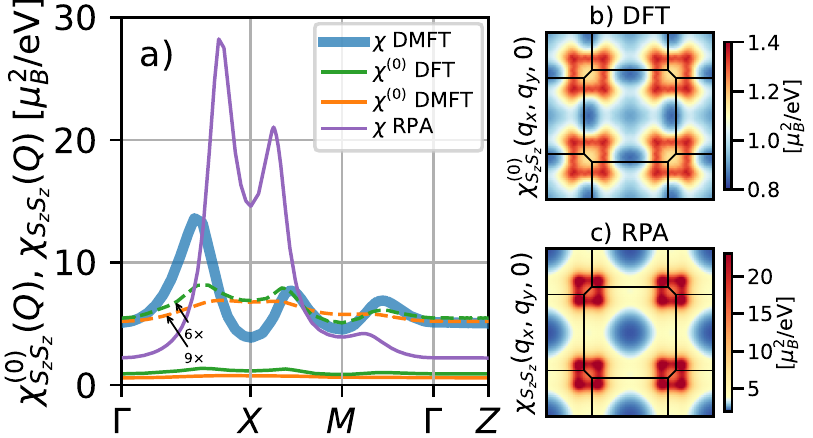}
\caption{\label{fig:methods}
  a) Spin-susceptibility $\chi_{S_z S_z}(\mathbf{Q})$ at $T=464\,$K on the high-symmetry path $\Gamma$--$X$--$M$--$\Gamma$--$Z$ (see Fig.\ \ref{fig:plane}). The DMFT response (blue) is compared to the the screened RPA result (purple) and
  the DFT (green) and DMFT (orange) bare bubbles $\chi^{(0)}_{S_z S_z}(\mathbf{Q}) \propto GG$. Note the scaling of the dashed lines.
  Planar cuts at $q_z=0$ for b) DFT and c) screened RPA are also shown, cf.\ DMFT in Fig.\ \ref{fig:plane}.
} \end{figure}

% --------------------------------------------------------------------
%\para{Levels of approximation:}
% --------------------------------------------------------------------
%
We now compare our results to simpler approximations
and show that the DMFT results are the only one qualitatively compatible with experiments
and that the full frequency-dependent vertex is a crucial part of the calculation which can not be neglected
\footnote{Note that a similar observation has recently been made for the charge response of the cuprates \cite{PhysRevB.99.035161}.}.
Indeed, in Fig.\ \ref{fig:methods}a the DMFT result is compared to the bare DFT and
DMFT bubbles ($\chi^{(0)}_\text{DFT}$, $\chi^{(0)}_\text{DMFT}$) and the screened RPA result $\chi_\text{RPA}$. 
The RPA calculation uses -- in spirit of Ref.\ \onlinecite{Gingras:2018aa} -- the DMFT bubble
$\chi^{(0)}$ and screened effective interaction parameters $\tilde{U} =
1.37\eV$ and $\tilde{J}/\tilde{U} = 0.4/2.3$, where $\tilde{U}$ has been taken
to reproduce the local susceptibility $\chi(\mathbf{r} = \mathbf{0}) \approx
7.3 \chiunit$ of DMFT.
The frequency dependent particle-hole vertex is clearly essential in the DMFT calculation, as
$\chi^{(0)}_\text{DMFT}$ is much smaller than the DMFT result $\chi_\text{DMFT}$.
$\chi^{(0)}_\text{DFT}$ is also strongly suppressed compared to $\chi_\text{DMFT}$, 
and the $X$-point response is higher than both the $\Gamma$ and $M$ points (see
Fig.\ \ref{fig:methods}b).
Finally, the screened RPA using the DMFT bubble $\chi_\text{RPA}$
severely overestimates the strength of the nesting peaks, underestimates the constant background response, 
and fails both to enhance $\chi(\QG)$ and to suppress $\chi(\QX)$, see Fig.\ \ref{fig:methods}c.

% --------------------------------------------------------------------
%\para{Conclusions \& outlook:}
% --------------------------------------------------------------------
%
In conclusion, we have analyzed the momentum-dependent magnetic response of
Sr$_2$RuO$_4$ using dynamical mean-field theory, taking full account of vertex
corrections. The latter are found to play a crucial role, leading to key
effects absent at the RPA level such as the suppression of the
antiferromagnetic response at $\QX$. 
In agreement with neutron scattering experiments \cite{PhysRevLett.122.047004},
the magnetic response has two main components: an SDW incommensurate response at $\QIC$ and a quasi-local weakly momentum-dependent component, which provides the main contribution to the overall momentum integrated response. 
Our main result, on a qualitative level, is the demonstration that the physical
origin of the quasi-local magnetic response is the Hund's coupling, hence
reconciling the experimental emphasis put on spin fluctuations in this material
with the theoretical picture of Sr$_2$RuO$_4$ as a `Hund's metal'. 

This has far-reaching consequences: both our theoretical calculations and neutron scattering experiments indicate that 
there is no dispersing `quasi-ferromagnetic' spin fluctuation mode in Sr$_2$RuO$_4$. 
Hence, pairing mechanisms based on a mediating bosonic mode (`glue') associated with ferromagnetic spin fluctuations \cite{0953-8984-7-47-002, PhysRevLett.79.733, PhysRevLett.82.4324, RevModPhys.75.657} have to be seriously reconsidered. The observed suppression of the magnetic response at the $X$ point also invalidates an antiferromagnetic `glue'. 
Instead, pairing mechanisms based on a quasi-local mode associated with Hund's
coupling offer a promising route.  Recent work has appeared in this direction
for model Hamiltonians \cite{PhysRevLett.115.247001, Hoshino:2016aa, Steiner:2016ab} and for iron-based superconductors \cite{PhysRevLett.121.187003}. However, these mechanisms were proposed in the regime of slow spin fluctuations above the Fermi liquid temperature, and need to be extended to be applicable to Sr$_2$RuO$_4$. This is a key agenda for future work aiming at solving the 25-years old puzzle of superconductivity in this material \cite{Mackenzie:2017aa}. 

% --------------------------------------------------------------------
% --------------------------------------------------------------------

% --------------------------------------------------------------------
\begin{acknowledgments}
The authors would like to acknowledge discussions with
L.\ Boehnke,
M.\ Braden,
X.\ Chen,
M.\ Ferrero,
A.\ Georgescu,
O.\ Gingras,
S.\ Hoshino,
Y.\ Maeno, 
I.\ Mazin, 
J.\ Mravlje, 
R.\ Nourafkan,
T.\ Schäfer,
A.\ M.\ Tremblay,
P.\ Werner,
and especially Y.\ Sidis.
AG acknowledges the support of the European Research Council (ERC-319286-QMAC).
AG and HURS acknowledge the support of the Swiss National Science Foundation (NCCR MARVEL) at the initial stage of this work.
The Flatiron Institute is a division of the Simons Foundation.
\end{acknowledgments}


% --------------------------------------------------------------------
\end{document}